**A Theoretical Solution of the Mind-Body Problem: An Operationalized Proof that no Purely Physical System Can Exhibit all the Properties of Human Consciousness**

Catherine M Reason[1]

*London, United Kingdom*

*This article presents an operationalized solution to the mind-body problem which relies on rigorously defined theoretical reasoning rather than philosophical argument. We identify a specific operation which is a necessary property of all healthy human conscious individuals -- specifically the operation of self-certainty, or the capacity of healthy conscious humans to "know" with certainty that they are conscious. This operation is shown to be inconsistent with the properties possible in any meaningful definition of a physical system. This inconsistency is demonstrated by proving a "no-go" theorem for any physical system capable of human logical reasoning, if this reasoning is required to be both sound and consistent. The proof of this theorem is both general -- it applies to any function whereby evidence affects the state of some physical system -- and recursive, since any physical process subserving a function of this type is shown to imply another such function. Thus for at least one aspect of human consciousness, the mind-body problem is now conclusively resolved.*

## Introduction

Of all the problems in the study of human consciousness, the mind-body problem has undoubtedly proven the most intractable to date. Chalmers (1996) has referred to this as the Hard Problem, of consciousness, and a study of the literature on this topic finds little agreement as to its likely solution. Some approaches (see for example Churchland, 1996; Dennett 1996; Hacker 2010) deny the problem even exists at all. So far, however, all approaches to the mind-body problem have been philosophical in character, and scientific analysis has tended to avoid the question. Chalmers' own approach to the problem relies on the notion of "philosophical zombies" -- beings physically identical to conscious human beings but without subjective experience. Chalmers' arguments are typically philosophical in nature, relying on concepts which are expressed in ambiguous verbal forms and which therefore provide opportunities for considerable misinterpretation.

The beginnings of a way out of this impasse came in 1995 when Gilbert Caplain published an outline for a proof that consciousness could not be a computational property using arguments drawn from computer science and logic, rather than philosophy. Caplain's proof was expressed only in outline, and necessarily relied on much philosophical terminology such as knowledge, justification and belief without giving precise theoretical definitions of these terms. However Caplain's proof did incorporate one highly innovative development -- it did not rely on some general, poorly defined notion of consciousness as an entity, but instead focused on one particular *property* associated with human consciousness. Furthermore the property Caplain chose, the capacity of human beings to know with certainty that they are conscious, is a property that can be represented as a specific *operation*. Thus Caplain pointed the way to an *operationalized* approach to the mind-body problem, in which undefined or poorly defined notions of consciousness or subjectivity could be replaced by rigorously defined *operations*. Caplain's proof therefore offered the promise of being expressible not as a philosophical argument, but as a logical theorem deriving from rigorously defined premises.

In a recent paper the present author (Reason 2016) attempted to expand Caplain's outline, by showing how it could be generalized to all physical systems and by attempting to remove all ambiguity arising from the use of undefined philosophical terms. Informal communications subsequently revealed this attempt to be only partially successful, and considerable ambiguity remained over the use of such terms as *certainty* and *reliability.* Two issues quickly became apparent from these informal communications. Firstly, that most readers were of the view that there must be something wrong with the proof; that it was in some way artificial, or contrived, or "playing with words". Secondly, that it was extremely difficult for some readers to disentangle their thinking from their intuitions and introspections about their own conscious experience. The only way of dealing with these difficulties is to present the proof in as rigorous a manner as I am able, even if the level of detail required for this sometimes looks like overkill.

The purpose of this current paper is threefold: Firstly, to explain the logic of Caplain's proof in an intuitive, easily accessible form; secondly, to illustrate how an operational approach can convert a verbal philosophical argument into a rigorous theoretical proof; and thirdly, to express Caplain's proof in just such a rigorous, theoretical form. Philosophers who read this paper are asked to bear in mind that there are important differences between a theoretical proof and a philosophical argument; many of the misconceptions which philosophers have had about this proof are the result of not fully appreciating these differences. Some of these misconceptions will be listed toward the end of this article.

**The rationale behind Caplain's proof**

The theoretical argument underlying Caplain's proof is extremely simple, and is in fact simply an extension of Descartes' evil demon thought experiment -- which tends to be presented in modern formulations as a "brain in a vat". For those unfamiliar with this thought experiment, here is a brief summary. We are asked to consider the possibility that we are in fact disembodied brains, kept alive in tanks or vats of nutrients, and that all our experiences of the outside world are in fact illusions created through direct stimulation of our sensory receptors by electrical or other means. The situation is in fact exactly that described in the well-known "Matrix" series of movies. The upshot of this is that our sensory experience is identical regardless of whether it is ultimately generated by the external world we think we perceive, or whether it is generated directly by means of some sort of deceptive apparatus. So we can never know for certain, on the basis of sensory experience, whether the external world really exists.

Caplain's proof simply extends this thought experiment from sensory perception to introspection. In a computing machine, the output of every act of either perception or introspection must be represented by some sort of symbolic state. This state must be the result of some computational process, and the only way of establishing the accuracy of that process is by the application of some other process. This leads to an infinite regress along the lines of Munchhausen's Trilemma.

The version of Caplain's proof which I am presenting here is constructed according to the following rationale. First, we operationalize consciousness in terms of an operation I shall call self-certainty, which will be described more fully in the next section. This enables us to represent consciousness as a simple function (the function of answering a YES/NO question). Self-certainty is thus a specific example of a general class of functions, consisting of all functions which answer YES/NO questions.

If we can then show that, for a given class of physical systems, no YES/NO question can be answered with certainty, we will have shown that self-certainty is impossible in that class of physical systems. The class of physical systems we choose is that class of physical systems which can exhibit humanlike logical reasoning, since human beings must by definition exhibit humanlike reasoning. Any system capable of humanlike logical reasoning we shall call $H$.

The proof which is to be presented here is the sequence of inferences in $H$ which establish that no physical system reasoning in $H$ can answer any YES/NO question with certainty. Therefore, if this proof A is sound, no physical system capable of reasoning in $H$ is capable of self-certainty.

It follows that, if we assume that all healthy conscious human beings are capable of self-certainty, we have shown that no healthy human being can be an exclusively physical system.

**Explaining the Operational Approach**

Consider the situation in this somewhat contrived little thought experiment:

Two philosophers are arguing about whether buses run on time in northwest London. Philosopher A wants to know whether the H12 bus runs on time. He decides to choose the Pinner High Street bus stop and counts how often the bus is on time at that particular stop. He finds the bus is late more often than not at that particular stop, and so concludes that the H12 bus does not run on time. So here we have:

Premise 1: If the H12 bus runs on time then it will stop at Pinner High Street on time;

Premise 2:  The H12 bus does not stop at Pinner High Street on time;

Conclusion:  The H12 does not run on time.

Clearly one can argue about the sampling difficulties associated with this little experiment, but let us leave that aside.  Instead we introduce Philosopher B, who examines the experiment and declares the first premise to be unsound.  According to Philosopher B, buses stop for all sorts of reasons; they may break down, or they may get stuck in traffic.  Or it may be a public holiday and the buses are not running at all.  It does not follow, says Philosopher B, that because buses do not stop at some particular place, and at some particular time, they must be running late.  Philosopher A, according to Philosopher B, has just made an unsound assumption about the meaning of the term "bus stop".

Of course it should be simple to work this out.  After all, everyone knows what a bus stop is and everyone should know what Philosopher A means by a bus stop.  But philosophy is prone to a rather peculiar disorder in situations like these, which I shall term a "sigh of relief and stop thinking" reaction.  By which I mean that Philosopher B has found an ambiguous meaning which he can misinterpret to his benefit, and so immediately sighs with relief and stops thinking.

Philosopher A now returns to his study and publishes an explanation of exactly what he means by a bus stop, and why his experiment depends only on that meaning of the term "bus stop",  and not on any other meaning.  But by this time Philosopher B has moved on.  He has completely forgotten the point of the original argument and is now making general assertions such as "There is no reason to suppose any relationship between the lateness of buses and the timing of their stops.  In fact the whole notion of a 'bus stop' probably lacks any coherent meaning to begin with."

The purpose of this admittedly rather absurd scenario is to illustrate a situation which will be familiar to anyone who has ever followed a philosophical debate; a situation which might be called "motivated misunderstanding".  Philosophical arguments are phrased in natural language terms, but their referents often involve metaphysical issues, or other issues which are hard to define unambiguously.  Chalmers (2011) refers to this type of problem as a "verbal dispute", and suggests various ways of resolving the difficulty.  However his methods of resolution depend on the assumption that both parties to a dispute are motivated to come to a common understanding.  One might suspect, given human nature, that some parties to a dispute might not have an incentive to arrive at a common understanding if the result is that their particular views are disproved or discredited.  Such individuals would therefore be motivated to prevent the ambiguity being resolved.  It would be difficult to arrive at a common understanding in such a case.

In science this situation is avoided, because scientific investigations try to avoid metaphysical questions and to concentrate on problems where the relevant factors can be strictly defined, observed and measured.  The operational approach is one way of doing this (Bridgeman 1927).  In the operational approach to measurement, for example, one avoids the temptation to regard measurements of physical quantities as "proxies" for some sort of metaphysical property. Consider the operation of measuring the length of some physical object by placing a measuring rod alongside it. One might be tempted to regard the reading from the measuring rod as a proxy of some sort for some metaphysical property called "length" which cannot be directly observed. However this raises inevitable questions about how the relationship between the proxy and the underlying metaphysical property can be established, given that the latter cannot be directly observed. These difficulties are circumvented if one regards the physical quantity "length" to be just the result obtained by placing a measuring rod alongside something. The output of the measuring operation is therefore to be taken as synonymous with the physical property "length". One can thus carry out measurements of physical quantities without worrying about what metaphysical properties, if any, may be associated with those quantities.

Of course there are limits to how far one can apply this approach to questions of interest to philosophers, since issues of metaphysics are fundamental to many of these questions. The nature of consciousness -- in particular the so-called "Hard Problem" of consciousness -- is a case in point. Caplain's work, however, shows that even in the study of consciousness, there are problems which can be treated by the operational approach and so lifted from the realm of philosophical speculation into the domain of theoretical and scientific analysis.

To see how this is so, we must note that Caplain's proof applies not to some indefinable metaphysical quality of consciousness, but to a very specific and well-defined *operation* -- that is, the operation whereby a conscious human being comes to "know" (in Caplain's words) with certainty that they are conscious. This operation can be expressed as the process of answering a binary question -- a question for which the answer is either YES or NO. If a conscious human subject asks the question "Am I certainly conscious?" , then there should be some operation associated with that subject's consciousness which is capable of giving the answer YES to that question. One can then ask whether that operation is one that can be performed by some physical system (although Caplain's original outline proof applied only to computing machines). If it turns out that the answer to this question is *no*, then the mind-body problem is at least partially resolved without any consideration having to be given to the metaphysical nature of consciousness itself. That is to say, one could prove that consciousness cannot wholly be a physical process without worrying about what consciousness itself actually is. This advantage of the operational approach saves us a lot of work. For example, one currently fashionable explanation for the Hard Problem is Illusionism (see for example Frankish 2016, Blackmore 2016). Illusionism is based on the premise that the properties of consciousness are not what they subjectively appear to be. But the operational approach enables us the sidestep the question of whether this doctrine is true, or even whether it is coherent.

Of course in order to do this one needs to do far more than just operationalize consciousness. One also needs to operationalize the concept of a physical system itself, and also define rigorously any other assumptions which are necessary to demonstrate an inconsistency between the operation of ascertaining that one is certainly conscious and the properties of physical systems. How might this be done?

If our intention is to demonstrate an inconsistency between an operationalized definition of consciousness and an operationalized definition of a physical system, then it is essential that any definitions and terms employed be unambiguous. Any terms associated with metaphysics, such as *belief,* or *knowledge,* or *justification,* should be avoided since these are likely to be interpreted in subtly different ways by different readers, and these different ways are likely to have hidden assumptions built into them. Likewise it is important not to get bogged down in the quagmire of human epistemology -- we are not interested in the question of how human beings "know" things, or "understand" things, or *how* they become certain of things. All such questions are irrelevant to our task and any terminology relating to such questions is to be avoided. To the extent we are interested in epistemology at all, we are interested only in the epistemology of physical systems.

**Assumptions necessary for an operationalized formulation of the proof**

In order to express the proof in a form that can reasonably be described as rigorous, it is necessary to list the assumptions and premises on which his proof depends, in a clear and unambiguous form. This we shall now proceed to do[2].

We first need some sort of definition of a physical system. It should be borne in mind here that the purpose of our definitions is to eliminate ambiguity, not to provide a comprehensive description. We can therefore operationalize the definition of a physical system in terms of some general property which a physical system must possess, without attempting to capture the full meaning of the term "physical system". We therefore define a physical system as some set of physical processes, where a physical process is defined as follows:

Definition: A *physical process* is some entity which has an objective existence and is capable of evolving in time.

Philosophers use the notion of *supervenience* to refer to entities which are not themselves physical systems but whose existence is nonetheless consistent with the doctrine of physicalism. An entity S is said to be *supervenient* on some basis B if two things which have the same B properties must have the same S properties (Kim 1984). Mental processes are thus assumed to supervene on physical processes. Hence the following expression:

*Principle of Physicalism*[3]: All mental processes supervene on some physical process or set of processes.

(The notion of a *mental process* will be expressed more formally in the next section.) Our operationalized notion of consciousness also needs to be expressed a little more specifically. We shall call this notion *self-certainty*:

Definition: *Self-certainty* is the capacity of at least some conscious beings to verify with absolute certainty that they are conscious -- that is, to give the answer YES to the question "Am I certainly conscious?"

It is important to emphasize that it is only *absolute* certainty which is at issue here. Caplain's proof does not imply that physical systems cannot be *reasonably* certain of being conscious, or that they might have a certainty of being conscious which is contingent on additional assumptions. For example Chalmers (2012) postulates an analogous situation involving mathematical reasoning; in this example an expert mathematician is given cause to doubt the accuracy of his own mathematical musings by being offered the possibility that he has been administered some sort of drug which destroys the capacity for mathematical reasoning. Chalmers suggests that the mathematician could regain confidence in the accuracy of his own cogitations by insulating himself from all such questions (that is to say, by ignoring them). Certainty of this sort, which depends on additional assumptions, is not covered by Caplain's proof and does not constitute self-certainty. We can express this as:

Condition 1: An entity is self-certain only if its certainty is absolute (beyond any possibility of error) and does not depend on additional assumptions.

It is also necessary to emphasize that self-certainty does not require assumptions to be made about the nature of consciousness. Any such assumptions -- for example, that consciousness equates to wakefulness, or that consciousness requires some specific sort of self-awareness -- should therefore not be made. A conscious being might, for example, be experiencing a lucid dream or some other altered state of consciousness. So long as the conscious being can verify that it is in some conscious state or other, the requirement for self-certainty is satisfied. We express this as:

Condition 2: Self-certainty does not require that the conscious state which is found to exist has any particular property or set of properties.[4]

Our next two premises refer to the rules of logical inference by which Caplain's proof can be established. Our objective is to prove Caplain's result as a rigorous theorem, but it must be admitted that this cannot be done in the context of any specified formal system. Even if such a formal system should exist, we do not know what it is. Fortunately, however, these details can be overlooked. If we start with the assumption that our own ability to reason logically, as human beings, is in principle sound[5] (which is a necessary assumption for any sort of theoretical work whatsoever)

---

[4]The significance of this condition cannot be overstated. Philosophers who find the operational approach objectionable (and this includes just about every philosopher who has ever contacted me on the subject) argue that nothing can be inferred about the nature of consciousness from such operational definitions. Condition 2 illustrates that this situation is intentional. The no-go theorem to be proven in the following section depends on this operational definition alone, and not on any philosophical notions about what consciousness is or should be. To underline this, one can express self-certainty as *illusory self-certainty*, as in the sentence "It is absolutely certain that I have at least the illusion of being conscious" without affecting the validity of the proof.

[5] At first glance this may seem to be inconsistent with Condition 1 -- however this is not the case. Condition 1 requires only that *self-certainty* does not require additional assumptions ; however it is necessary to make additional assumptions in order to go beyond self-certainty and to prove Caplain's result.

then we can conclude that the necessary rules of inference are precisely those which enable us, as human beings, to reason logically. Any system which incorporates these particular rules of inference, we shall call $H$. This leads to the following two assumptions:

Assumption 1:  There exists a physical system M which supports the logical system $H$;

Assumption 2:  $H$ is in principle both sound and consistent.

"Sound in principle" simply means that if a system reasoning with $H$ makes a mistake, there is no reason in principle why it should eventually not correct that mistake.[6] $H$ does not therefore condemn any system using it to fallacious logical reasoning which can never be corrected. By "correct", we mean of course that $H$ incorporates a capacity for classical inference. There may well be many different formal systems, with different sets of axioms and different rules of inference, which exhibit the necessary properties of $H$, or it may be that there is no such formal system. By defining $H$ in the way we do, we manage to avoid this question entirely. However since $H$ is defined only implicitly, derivations in $H$ cannot be given in the way they would be for formally specified systems. Such derivations must be presented instead in terms of the intuitive logic which supervenes on $H$. For this reason the following proof must be presented in somewhat tedious detail, in order that each inference may be checked individually by the reader.

### An operational formulation of Caplain's proof

We are now in a position to present Caplain's proof in a rigorous form. The proof will now be presented in a number of stages, and each milestone will be clearly identified and numbered. The reader is invited to consider the reasonableness of each inference as it is presented. Should any reader find the accompanying profusion of Greek letters and other algebraic symbols somewhat bewildering, they may find it useful to consult the simpler proof in Reason (2016), in which they do not appear.

It will be convenient to restrict our considerations to mental processes which can be represented as functions. We will define M to be some physical system, and $\varphi$ to be some function which yields the answer to a YES/NO question. We shall begin with the simplest case in which all questions can be considered to refer to propositions which are either TRUE or UNTRUE. The function $\varphi$ therefore represents the mapping

TRUE $\rightarrow$ YES

UNTRUE$\rightarrow$ NO

---

[6] Another way of looking at this, which is perhaps strictly more accurate, is to assert that $H$ is sound but that individual intelligent beings are noisy theorem-provers in $H$. That is to say, intelligent beings can prove theorems in $H$ but are subject to random errors. However they are not subject to *systematic* errors.

where TRUE and UNTRUE are the possible truth-values of some proposition p. We shall call the range of this function the *evidence state* of p. So if T represents the truth-value of p, and K represents the evidence state of p, the mapping can be represented as

$$K = \varphi(T)$$

We now present this as our first milestone:

Milestone 1: Any mental process which is equivalent to the process of answering a YES/NO question can be expressed as a function of the form $\varphi$.

The equivalence covers mental processes which may not themselves be answers to YES/NO questions, but can be represented as such: For example, the mental process of thinking "I exist" is equivalent to the process of answering the YES/NO question "Do I exist?"

In a physical system M, an evidence state of some proposition p can be any state of M correlated with the correct truth-value for p, and the function $\varphi$ will be performed by some physical process P, as is required by the Principle of Physicalism. This gives us a general format for mental processes in a physical system. If T is the truth value for some proposition p, and K is the corresponding evidence state for T, then

$$K = O(P)$$

where O is the output of some physical process P. This is to say that if TRUE is the correct truth value for p, then M will evolve to state $K_T$, and if TRUE is not the correct truth value for p, then M will evolve to some state $nonK_T$. Thus P executes the mapping:

$$TRUE \rightarrow K_T$$

$$not\ TRUE \rightarrow nonK_T$$

where $K_T$ and $nonK_T$ are separate states of M. We shall call this mapping $\pi$.

It is possible that T is a physical state of M, and furthermore that T is the same physical state as K; $K_T$ however cannot be the same physical state as $nonK_T$. Some decision is therefore always required as to whether M evolves to $K_T$ or to $nonK_T$. This decision , and the subsequent evolution of M, constitute the process P. This leads us to our second milestone:

Milestone 2: Any physical instantiation of a function in the form of $\varphi$ can be expressed as a mapping in the form of $\pi$, which will be performed by some physical process P.

Now consider the mapping which does the *reverse* of $\pi$, that is the mapping:

TRUE → nonK$_T$

not TRUE → K$_T$

a mapping which we shall call ρ.  It is nonetheless clear that ρ has the same *form* as π, and by Milestone 2 there will be some physical process which performs it, which we shall call R.  We shall express this as our next milestone:

Milestone 3:  For every mapping of the form π performed by a physical process P, there will be some contradictory mapping of the form ρ and some physical process R which performs this[7].

But if K represents the evidence state of some true proposition, how is M to ascertain whether the state K has been produced by P or by R?  It clearly matters, because if K has been produced by P then K$_T$ will represent the answer YES to some question, but if K has been produced by R, then K$_T$ will represent the answer NO to that same question.   To resolve this difficulty, it is clearly necessary for M to ascertain whether K has been produced by P or by R!

Another way of looking at this is to say that P is an *accurate* instantiation of φ but that R is an *inaccurate* instantiation of φ.   In this case K$_T$ will always represent the evidence state YES, but if K has been produced by P then K will be a correct evidence state, whereas if K has been produced by R then K will be an incorrect evidence state. This brings us to the notion of *failure*, which is always a necessary consideration when dealing with processes in the real world, as opposed to the idealized mathematical functions they perform.  We can express this notion in terms of the following axiom:[8]

Axiom of Fallibility:  Given any arbitrarily selected physical process, whose properties have not been ascertained, it is impossible to be certain that this process will or will not perform some given function.

In other words, if we have a physical system M in which there exists some evidence state K, produced by some process which is supposed to instantiate a function of the form φ, it is impossible to be sure whether that state has been produced by the process P or by the process R without ascertaining any of the properties of the process.   We might be able to deduce through logical deduction what that process should be, but we cannot through logical deduction alone deduce what that process actually *is* in any given case.

How could we ascertain the properties of this mysterious process?  We could either examine the process directly, or we could examine the mapping performed by the process, which entails examining the states between which the mapping occurs.  In the former case we could ask, for example, "Does this process have the properties of P or R?"  This is clearly equivalent to asking either the questions "Does this process have the properties of P?" and "Does this process have the properties of R?"  These are clearly YES/NO questions and so by Milestone 1 can be expressed as functions of the form φ -- let us call this particular function $\varphi_1$.

---

[7]   This can always be done by applying a logical NOT operator to the truth value before performing φ ☉.  Therefore if P exists, so does R.

[8]   It is convenient at this point to express the notion of failure as an axiom of *H*, but it is not strictly necessary.  Later we shall show how it is possible to express the notion of failure without assuming the Axiom of Fallibility.

In the second case we could ask, for example, "Does this process map T on to K?"  This is also clearly a YES/NO question and so is also expressible as a function of the form φ.  Let us call this particular function $\varphi_2$.

By Milestone 2, any physical instantiation of these functions can be expressed as a mapping of the form π, which will be performed by some physical process.  In either case call this process P'.

By Milestone 3 there will exist for this mapping a corresponding mapping of the form ρ, which will be performed by some physical process we shall call R'.

From the Axiom of Fallibility it follows that one cannot determine if the *actual* physical process which instantiates $\varphi_1$ or $\varphi_2$ in the physical system M, is P' or R' (or indeed some completely different third process as yet unidentified) without ascertaining the properties of this actual process.  But ascertaining the properties of an actual physical process (as opposed to an idealized one) is equivalent to a function of the form φ.  And by the Principle of Physicalism, M can only perform functions of the form φ by means of some physical process.  This leads to the following rather awkward state of affairs:

A physical system M can only perform some function of the form φ by means of some physical process;

M can only ascertain the correctness of some physical process by performing some function of the form φ.

This is either circular or leads to an infinite regress.  The nature of the infinite regress is easily demonstrated by examining the functions which need to be performed.  If the first function can be represented as

$$K_0 = \varphi_0(T_0)$$

where $K_0$ is the evidence state of some proposition $p_0$ and $T_0$ is the truth value of $p_0$; then subsequent functions can be represented as

$$K_i = \varphi_i(T_i)$$

where $K_i$ and $T_i$ are the evidence state and truth value respectively of some proposition $p_i$ where $p_i$ is equivalent to answering the YES/NO question:

"Is the function which answers the question equivalent to $p_{i-1}$ correctly performed?"

Clearly the number of questions, which would need to be answered in order to establish that any given φ-type function has been performed correctly, is at least countably infinite.

The upshot of all this is that M can never establish with provable accuracy whether any of its processes is accurate or inaccurate[9], since establishing this requires a non-terminating sequence of processes. It does not yet follow, however, that M cannot arrive at the conclusion that all its processes are accurate. M might, indeed, be correct to conclude that all of its processes are accurate. However such a conclusion cannot be a logically valid theorem in $H$; there is no logically sound chain of reasoning in $H$ by which M can show that any of its processes is accurate. To put this in slightly less technical language, it is impossible to prove in $H$ that any of M's processes is accurate.

If this is not yet obvious, consider the following: We have established above a chain of reasoning which shows that any attempt by M to establish the accuracy of any of its processes logically implies an infinite regress. We have established this conclusion by means of our own human capacity for logical reasoning; a capacity which we have labeled $H$. By Assumption 2, $H$ is in principle sound and consistent. If $H$ is sound and consistent, then it cannot support a stable[10] chain of reasoning which would lead to a conclusion inconsistent with other conclusions already established in $H$; that is to say, with other theorems in $H$. M might be able to derive a logically valid proof in some other system, say $nonH$, that some of its processes were accurate; but then by definition M could not be a human being.[11] One can therefore say that the statement "M cannot prove the accuracy of any of its processes" is equivalent to a theorem in $H$.

If at this point we invoke Assumption 1, then M itself can establish that none of its processes is provably accurate. Since M can reason according to the system $H$, M can prove any theorem that can be proven by a human being; this includes the proposition "M cannot prove the accuracy of any of its processes". To express this in slightly less formal terms, we can say that if a human being, reasoning according to the principles of $H$, can show that M cannot establish the accuracy of any of its own processes, then M can do the same.

This is an extremely raw and primitive version of the proof we need, which applies only to propositions which are either categorically true or categorically untrue. In real

---

[9]It would strictly be more accurate to say that the accuracy of M cannot be proven in the logical system $H$.

[10]Since $H$ is only sound and consistent in principle, it could temporarily support an invalid chain of reasoning inconsistent with other theorems in $H$. However such a chain of reasoning would have to be unstable, by which I mean that $H$ would also have to support a chain of reasoning showing this fallacious chain to be invalid. See also Note 5.

[11] Such a system might, for example, contain axioms which simply assert that certain physical processes are accurate. It would be necessary for such a system to be so constituted that these additional axioms did not lead to contradictions with other properties of the system; this would entail $nonH$ abandoning certain axioms, such as the Axiom of Fallibility, and having different rules of inference from $H$. The rules of $H$, for example, allow M to ask the question "Is there any reason why these additional axioms should not be dropped?"

life, we are often concerned with degrees of probability as much as simple truth or untruth. Often we will have to deal with questions where we simply do not know the answer; where the possible answers to a YES/NO question will be YES, NO and MAYBE. This is easily remedied, however, by expressing such questions in the form:

"Is it the case that some proposition p is certainly true?"

To avoid ambiguity we shall define any statement of the form "p is certainly true" as meaning that there is zero probability of p being untrue, or that there exists no possible world in which p is untrue. Statements of the form "M is certain of p" will be defined as meaning "M is in the evidence state YES for some particular proposition p".

All questions of this type can be expressed as YES/NO questions. We can now encapsulate the reasoning in this section as a specific proof, which we shall call Proof A. This proof establishes that no physical system which possesses reasoning capacities equivalent to those of a human being can correctly answer YES to the following question:

"Am I certain that I am answering some given YES/NO question correctly?"

We can express this as our next milestone:

Milestone 4: For any suitably qualified physical system M (that is, one which is capable of human-level reasoning), and for any proposition equivalent to the answer to a YES/NO question, there can exist some physical process of type R which will answer that question inaccurately. Therefore any suitably qualified physical system can deduce that it can never be certain that its answer to any YES/NO question is correct.

This is an extremely general result and it is thus worthwhile checking for any possible loopholes. The reader might, for example, consider that since Proof A applies only to physical systems, it might not apply to M if M is not certain that it is a physical system. We can deal with this by expressing Milestone 4 in a conditional fashion:

"If M is a physical system there can always exist some physical process of type R which will answer any given YES/NO question inaccurately."

Clearly this statement is provable in *H*. In order to establish that Proof A certainly does not apply to M, M must be able to answer "YES to the question:

"Is it certainly true that M is not a physical system?"

We will call the proposition equivalent to this question $p_0$. M can now ask the question $p_1$:

"Is there some process $R_0$ in M that would answer $p_0$ incorrectly?"

Whatever the answer to $p_1$, M can now ask the question $p_2$:

"Is there some process $R_1$ in M that would answer $p_1$ incorrectly?"

Clearly we find ourselves in another infinite regress. In fact generally for any proposition $p_i$ which can be expressed as a YES/NO question, there will be some proposition $p_{i+1}$ of the form:

"Is there some process $R_i$ in M that would answer $p_i$ incorrectly?"

Thus the sequence of questions which need to be answered in order to establish the truth of the proposition

"M is certainly not a physical system"

is non-terminating. Since M is in fact a physical system, each such question is equivalent to a φ-type function which must be performed by some physical process, which means the sequence of such processes is also non-terminating. So long as M is in fact a physical system, it does not matter if M itself is not certain that it is a physical system.

Having established this, our first application of Proof A is to the question

"Is it certain that I am conscious?"

where this question is subject to Conditions 1 and 2 as described in then previous section. Any healthy, conscious human being should be able to answer YES to this question. Proof A, however, shows that no suitably qualified physical system can do so. Conscious human beings, therefore, cannot be exclusively physical systems.

Before proceeding to the next section, we will now deal with the outstanding matter of the Axiom of Fallibility. The purpose of this axiom is to express the concept of failure, by asserting that there is no way of establishing through purely logical means whether some particular physical process is working correctly or incorrectly. Some readers, however, and most especially philosophers, may be attracted to the idea of simply dropping this axiom. One can however deal with the notion of failure without expressing it explicitly as an axiom. Consider some function of the form φ which is performed by some physical process. M can now ask the question

"Is it certain that the process which was supposed to perform the function φ did not in fact perform a mapping of the form ρ?"

to which the answer must be YES. The process of answering this question is itself a function of the form φ -- call this φ'. By the Principle of Physicalism, any physical system which has performed this function must have done so via some physical process. M can now ask the question

"Is it certain that the process which was supposed to perform the function φ' did not in fact perform a mapping of the form ρ?"

to which the answer must be YES. Since the process of answering this question is also a function of the form φ, which we can call φ'', it must by the Principle of Physicalism have been performed by some physical process.. In fact in general any function $\varphi_i$ which is assumed to have been performed correctly implies the existence of another function $\varphi_{i+1}$ which answers the question equivalent to that assumption. This implies a non-terminating sequence of functions and a corresponding non-terminating sequence of physical processes to perform them. There is, in other words, no escape from the infinite regress by abandoning the Axiom of Fallibility.

**Verbal summary of the proof**

Proof A has two important properties. Firstly, it is *generalizable*; it applies to any mapping of the form φ. Secondly, it is *recursive*; it shows that the question of whether any given φ-type function is performed correctly can itself be represented as a φ-type function, to which Proof A applies.

We shall now summarize the proof given in the last section in a somewhat more accessible, verbal form. In doing so we shall illustrate the properties of generality and recursiveness, and show how these can be used to deal with the apparent objections to the proof which have been raised informally by some philosophers. It will be useful to base our verbal summary on the notion of *evidence*; however we shall ignore all considerations about the nature of evidence and instead operationalize the *function* of evidence as a sequence of YES/NO questions. That is, if M has evidence that some proposition is true, then it is the function of that evidence to answer the question "Is this proposition certainly true?" with either a YES or a NO.

This function we shall refer to as the *evidence function*, which relates the presence or absence of some evidence to the evidence state K of some proposition. If we define a state E so that E is 1 when some evidence exists, and E is 0 when that evidence does not exist, then the evidence function relates E to K as follows:

If E = 1 then K = YES

If E = 0 then K = NO

By the Principle of Physicalism, any evidence function in M must be performed by a process which supervenes on some physical process or other. Since we have so far ascertained nothing about this process, we shall call it X.

Philosophers, and most particularly epistemologists, traditionally use possible worlds semantics to describe conditions of possibility and necessity. For convenience we shall adhere to this convention henceforth; if some proposition is possibly true, we shall express this by saying there exists some possible world in which that proposition is true. We shall refer to a possible world as *epistemically available* to M if M cannot rule out the possibility that M exists in such a world, given the evidence currently available to M. We shall say that M is certain of some proposition if M can correctly decide that there is no possible world *epistemically available* to M, in which the evidence state which corresponds to the truth of that proposition is itself incorrect.

We shall express the first part of our proof as the following logical argument, which we shall call Castor:

It follows from the Principle of Physicalism that any evidence function must supervene on some objectively real process X. Since X is an objectively real process it cannot be guaranteed a priori to be a correct instantiation of any function. We shall say that X *fails* on any occasion on which it does not correctly map some element from the domain of the function to its range. If X fails than its output will be an incorrect evidence state.

How can M be certain that the evidence state K is correct, ie that the mapping from E to K has been correctly performed? Only by ruling out the possibility that X has failed. In other worlds, there must be no possible world epistemically available to M in which X has failed. The question of whether X has failed can be represented as the question "Is it certainly the case that X has not failed?" The process of answering this question is itself equivalent to a new evidence function.

This argument is general -- it applies to any evidence function. We can express this as the second part of our proof, which we shall call Pollux:

Castor has generality, ie it applies to any evidence function. Therefore Castor can be applied recursively to the new evidence function generated by Castor. This will generate a third evidence function, to which Castor can be applied recursively yet again. Indeed each time Castor is applied to any evidence function, it will generate a another evidence function to which Castor can be applied. Therefore the correctness of any objectively real process X cannot be ascertained without performing a non-terminating sequence of functions, each of which (by the Principle of Physicalism) must be performed by some objectively real process.

Castor and Pollux together are equivalent to the Proof A given in the previous section. By splitting the proof into two sections in this way, the reader can clearly see how Proof A exhibits the two useful properties of generality and recursiveness. Castor establishes generality; Pollux shows that Castor can be applied recursively.

The foregoing tells us that M can never guarantee its own correctness when evaluating the truth of any proposition.. However this result has not been proven by M, but by us, the readers. It has been proven in whatever logical system or systems underly our reasoning processes, which we must assume are sound in principle. As in the previous section, let us call this system *H*. However Assumption 1 allows M to "inherit" *H*, as it were, and so to prove Proof A for itself. This reasoning also allows us to disregard the detailed character of *H* itself, since these details effectively cancel out. The reader's understanding of Proof A can effectively be regarded as a derivation of Proof A in the logical system *H*.

We shall now express the outcome of Proof A as a theorem, which we can call *Gemini*:

No suitably qualified physical system can exhibit the property of self-certainty, and in any system which exhibits self-certainty, the process which subserves self-certainty cannot depend on any objectively real process.

Readers should note that Proof A is a proof of exactly this theorem, and not any other. A former reviewer of this paper attempted to paraphrase this theorem into a generic and rather ambiguous statement about the epistemology of consciousness, and then expressed objections based on the resulting ambiguities. Such a practice is strongly to be discouraged. A theoretical proof is a proof of a specific, well-defined proposition, derived from a specific set of assumptions according to clear rules of inference. In this respect a theoretical proof is quite different from a philosophical argument.

Another common objection from philosophers is that conscious entities do not need to calculate or prove that they are conscious, but can in some way simply "ascertain" it. The mistake here is to assume that physical systems must necessarily have the same properties as conscious human beings. Any such "ascertaining" can still be represented as an evidence function, and if such an evidence function is performed by some physical process X (as is required by the Principle of Physicalism) then Proof A will apply to it. Readers should be careful not to attribute unconditionally to physical systems the properties of their own consciousness. Proof A applies generally to any system capable of reasoning in $H$, if both the Principle of Physicalism and the Axiom of Fallibility are assumed.

Another objection which has been made frequently to me in private communications is that it would be a trivial matter to program a machine to answer "Yes" whenever the question "Are you certainly conscious?" was asked. Of course this is true, but such a machine would, by definition, not be reasoning according to the rules of $H$. Therefore such an objection has no relevance to the Gemini theorem.

It has also been suggested that the correctness of X does not have to be established for M to be certain that the evidence state K is correct,[12] since in the possible world where X is correct, M may have access to other, different evidence from the possible world in which X fails. However this argument implicitly assumes another evidence function, and Proof A will apply recursively to any such function. One can express this evidence function in terms of the question "Does this other, different evidence demonstrate the truth of some proposition?" Since Proof A is generalizable to all evidence functions, it will apply to this one  This is true regardless of what evidence M may have access to, since evaluating any such evidence is equivalent to the question "Does this evidence demonstrate the truth of some proposition?" and the function of answering such a question will always be subject to Proof A. No matter what evidence M may have access to, therefore, M can never be certain that the evidence function implied by that evidence has been correctly performed.

Some philosophers insist that any notion of certainty entailing an infinite sequence of propositions is epistemologically unacceptable and unnecessary, and that certainty requires only that M is certain of some proposition p. This would be a reasonable remark if the notion of certainty employed here were axiomatic, but for physical systems reasoning in *H* this is not the case. The requirement for an infinite sequence of functions is in fact a therorem in *H*, the proof of which is implicit in Proof A. This proof can be expressed explicitly as follows: We start by assuming that M is certain of some proposition p. By the Principle of Physicalism, the process of determining p must supervene on some physical process, say X. From the Axiom of Fallibility, M can deduce that X may be fallible. By Assumption 1, M can deduce that if X is fallible, then the evidence state for p may be incorrect, in which case M cannot be certain of p. Therefore to be certain of p, M must have some way of determining that X is operating correctly. By the Principle of Physicalism, this must supervene on some physical process, say X*. But what process is X*? It cannot be X, since M relies on the process X* to ensure that X is operating correctly. X therefore depends on X*, not the other way round.

---

[12] I am indebted to David Chalmers for this objection, and also for the one in the subsequent paragraph.

Therefore M must assume that X* is some new process X', which is different from X. But from the Axiom of Fallibility, M can deduce that X' may be fallible. Since M's certainty of p depends on X, and the correctness of X depends on X', M must have some means of determining that X' is operating correctly. Let us call this process X**. But what process is X**? It cannot be X', since the correctness of X' depends on the correctness of X**. Neither can it be X, since establishing the correctness of X requires the correctness of X'. Therefore we require some new process X" ... and so on ad infinitum. This leaves us with no choice but to discard the assumption which led to the regress -- that is, the Principle of Physicalism.

A rather more serious objection to the Gemini proof is that the rules of *H* by which the proof is derived are never explicitly defined. The proof is derived using the rules of classical logic and we must therefore assume, firstly, that our ability as human beings to use these rules is sound, and secondly, that our judgement that these rules are the correct ones to apply in this situation is also sound. It is this second judgement which is impossible to formalize. Somehow, *H* must allow us (and any system which reasons like us) to decide that it is classical logic, rather than, say, some paraconsistent or quantum logic, which is the correct logic to use in this case. I do not believe that this is a serious problem, but it does underline how the proof depends on accepting Assumption 2.

A brief mention must be made of two alternative philosophical approaches which some people appear to believe constitute loopholes in Proof A. The first is *coherentism* -- the doctrine that although individual processes may not be reliable, a large ensemble of processes together might be. This is obviously vulnerable to applying Proof A simultaneously to every process in the ensemble. Alternatively, the ensemble can simply be treated as a single process. More fundamentally, however, coherentism cannot be proven to be true; it must therefore be assumed to be true, which entails adding another axiom to *H*. Since the rules of *H* must allow M to ask the question "Is there any reason why this axiom should not be dropped?" and since the answer to this question is by definition NO, coherentism cannot establish absolute certainty.

The second philosophical approach which requires some mention is *constitutivism* (Shoemaker, 1990). The key principle here is that the evidence state for the question "Am I certainly conscious?" is a physical state which overlaps wholly or partly with the physical basis of consciousness itself. This idea was dealt with briefly in Reason (2016); the problem here is that, while self-certainty might be allowable if constitutivism is true, the process of determining if constitutivism is true can itself be represented as an evidence function, to which Proof A will of course apply. More specifically, for constitutivism to work, M must (by the rules of *H*) be able to answer YES to the question "Is the evidence state for consciousness physically identical with consciousness itself?" This question clearly implies an evidence function, to which Proof A applies.

Proof A can be expressed in as a couplet of statements, as follows. For any suitably qualified physical system:

Any evidence function must, by the Principle of Physicalism, be performed by some physical process.

The correctness of any physical process must, by the Axiom of Fallibility, be determined by some evidence function.

All philosophical objections to Proof A can be dealt with by applying this *Gemini couplet*.

### What kind of process is not a physical process?

The reader may well now be wondering, if Proof A is as generalizable as it appears to be, why it does not simply apply to *all* processes, whether physical or not. What sort of characteristic could possibly differentiate physical processes, which are subject to Proof A, from non-physical processes, which are not subject to Proof A?

The answer is that physical processes are defined as being objectively real -- that is, they are assumed to have properties which exist independently of the subjective states which record the values of those properties. Therefore if we assume a φ-type function performed by some objectively real process X, there will be an objectively real fact about whether X is a process of type P, or a process of type R. The infinite regress, in the form of the non-terminating sequence of processes generated by Proof A, arises from the need to identify exactly what this objective property is. No such difficulty arises if no objectively real process exists, since there is then no objective property to be identified. The consequence of this however is that whatever process performs the φ-type function equivalent to self-certainty, this process cannot depend on any objectively real process -- it cannot, in other words, depend on any process which is external to human consciousness. Another way of looking at this is to say that any process which performs self-certainty must be subjectively real but not objectively real.

### Discussion

The Gemini theorem itself shows only that self-certainty is impossible in any physical system. In order to apply this to the mind-body problem we need to make a further assumption, which is that self-certainty is in fact possible for human beings. This is by its nature an empirical question. Some will regard it as intuitively obvious that human beings are capable of self-certainty, given that self-certainty does not require us to make any assumptions about the nature of consciousness or the properties of conscious states[13]. For those who do not, however, there is a clear empirical prediction which can be made.

Gemini implies that self-certainty will entail a violation of the conservation of energy (as was explained in Reason, 2016). We can express this violation as an inequality between the energy produced in a physical system and the energy consumed by that system. The difference between energy produced and energy consumed is denoted by the symbol $\chi$ (Reason 2016). It is this inequality which we must somehow detect experimentally.

One should bear in mind here that there is no a priori reason why $\chi$ should be a single event; it is quite possible the $\chi$ effect in the brain is made up of many separate, localized effects which all contribute toward a change in the brain's physical evolution. If one denotes these local effects by $\chi_n$, then the total effect $\chi$ will be given as follows:

$$\chi = \chi_1 + \chi_2 + \ldots\ldots + \chi_n$$

Since some of these local effects will be negative, it is theoretically possible for the $\chi$ effect over the whole brain to sum to zero. This would imply an ad hoc degree of symmetry in the brain's functional architecture which seems to me unlikely. Nonetheless we are more likely to be able to detect non-zero $\chi$ effects in highly localized regions of the brain than in larger, more general regions.

In order to do this we need to rely on proxy measures of energy production such as cerebral blood flow, and proxy measures of energy consumption such as changes in the brain's electromagnetic field. If $E_p$ represents some proxy measure of energy production, and $E_c$ some proxy for energy consumption, then the absolute $\chi$ value in some small three-dimensional compartment of the brain will be:

[13]A special problem exists in the case of self-referential statements such as the question "Is it certain that I exist?" In such a case a system reasoning in *H* could get as far as Milestone 2 and then argue, along the lines of Descartes' *Cogito ergo sum*, that it does not really matter if the physical process which subserves the answering of that question is accurate or not -- the mere fact that such a physical process exists is enough to answer the question in the affirmative. One can express this in the form of a *Cartesian Principle*: If P exists then M exists (since P is a process in M).

Descartes' *Cogito* can thus be represented in *H* as a syllogism of the form:

If P exists then M exists;
P exists;
Therefore M exists.

However this syllogism requires M to establish that it is the case that P exists. Since this clearly entails a φ-type function, Proof A will apply to it, as before.

$$\chi_{xyz} = |((f(E_p) + \text{energy}_{in}) - (g(E_c) + \text{energy}_{out}))|$$

where $\text{energy}_{in}$ and $\text{energy}_{out}$ are terms representing the energy flowing into and out of the compartment from other compartments. The Gemini theorem predicts that this quantity, summed over three dimensions, will be non-zero. In practice, measuring the energy transfer between compartments is not practical. The result of this is that only strongly asymmetric total $\chi$ effects (those that do not sum to zero) are likely to be detectable. Such asymmetric effects will be indicated by a residual effect $\chi_{res}$ given by:

$$\chi_{res} = f(E_p) - g(E_c)$$

when this is summed over three dimensions. The detection of symmetric $\chi$ effects, however, relies on assuming that the $\text{energy}_{in}$ and $\text{energy}_{out}$ terms can both be set to zero. Whether this is possible is unclear at the present time..

What sort of proxy measures of energy production and energy consumption are possible? The first problem to be confronted is that *in vivo* neuroimaging techniques assume the conservation of energy, and therefore do not explicitly distinguish between energy production and consumption, regarding both as proxy measures for neural activity. Obviously we cannot make this assumption. We therefore need to eaxmine possible neuroimaging techniques a little more closely.

The most widespread imaging technology used in neuroscientific research is Functional Magnetic Resonance Imaging (fMRI). This relies on the differing magnetic properties of oxygenated versus deoxygenated blood (Huettel, Song and McCarthy, 2014). Increased neural activity is thought to trigger a release in blood oxygenation levels, which is detectable in the form of the Blood Oxygenation level Dependent (BOLD) signal. Since there is a timelag between the increase in neural activity and the resulting increase in blood oxygenation, the temporal resolution of this technique is limited to about five seconds. Spatial resolution for this technique is less than five millimeters. This may be sufficient for detecting symmetric $\chi$ but it is not clear if intercompartmental energy transfer can be ignored using this method.

A more direct measure of energy production is Positron Emission Tomography. This can be used to measure glucose uptake by use of the radioactive tracer fluorodeoxyglucose (FDG). (See Vanitha 2011 for a brief description of the technique.) However the requirement for a radioactive tracer, together with the complexity and expense of the equipment, make this method problematic for pure research purposes. A possibly more promising approach is Phosphorus Magnetic Resonance Spectroscopy (P MRS) which measures the energy generated from the hydrolysis of ATP (Zhu, Qiao, Du, Xiong, Liu, Zhang, Ugurbil and Chen, 2012).

Measuring the other side of the problem, energy consumption, turns out to be a little more constricted. The most obvious technology is Magnetoencephalography (MEG), a description of which can be found in Hamalainen, Hari, Ilmoniem, Knuutila and Lounasmaa (1993). MEG measures neural activity in the form of magnetic fields generated by dendritic currents. Since it requires at least 50,000 neurons to generate a field strong enough to be detectable, this technique has limited spatial resolution (>

3mm) which depends on making assumptions about the nature of the neural activity which has generated the signal; however it has excellent temporal resolution.  A less well-known technique is Event-Related Optical Signal detection (Gratton and Fabiani 2001).  This method detects neuronal activity by via selective absorption of infra-red light.  It has excellent temporal resolution but its spatial resolution is limited to about a centimeter.  It also has the disadvantage that the technique only works to a depth of up to 5cm into the head, and it is not clear if $\chi$ can be assumed to occur within the cortex.  However a potential advantage of EROS is that it can be carried out simultaneously with a measurement of hemodynamic response in the form of NIRS, or  Near Infra-Red Signal detection (Gratton, Goodman-Wood and Fabiani 2001).  This property could be especially useful in detecting $\chi$.

It is the my hope that some or all of these techniques may offer the basis for a practical experiment to detect the $\chi$ effect.  Anyone who believes they can assist in carrying out such an experiment is urged to contact the author.